\documentclass[oneside,10pt]{swissi}

\newcommand{\papertitle}{Multi-Agent AI Architecture for Regulated Insurers}
\newcommand{\papersubtitle}{A generic AI framework under Solvency II and the AI Act in Austria and Germany}

\newcommand{\papercitationauthors}{Kurz, W.}

\newcommand{\paperauthors}{%
  Walter Kurz\textsuperscript{1}%
}

\newcommand{\paperaffiliations}{%
  \textsuperscript{1}Swissi Institute for AI,
  \href{mailto:kurz@swissi-ai.institute}{kurz@swissi-ai.institute}%
}

\newcommand{\paperdeclarationaffiliations}{%
  \begin{tabular}[t]{@{}l@{}}
    \textsuperscript{1}Swissi Institute for AI,
    \href{mailto:kurz@swissi-ai.institute}{kurz@swissi-ai.institute}%
  \end{tabular}%
}

\newcommand{\papercorresponding}{%
  Walter Kurz,
  \href{mailto:kurz@swissi-ai.institute}{kurz@swissi-ai.institute},
  \href{https://orcid.org/0009-0006-8045-4775}{ORCID 0009-0006-8045-4775}%
}

\newcommand{\paperfunding}{%
  This research received no external funding.%
}
\newcommand{\paperconflictsofinterest}{%
  The author declares no conflicts of interest.%
}
\newcommand{\paperdataavailability}{%
  The data and code supporting this article are available from the corresponding author on reasonable request.%
}
\newcommand{\paperaitools}{%
  Generative AI tools were used for editorial work only, such as language editing and formatting, in accordance with international scientific standards. The author verified the content and remains fully responsible for the article.%
}
\newcommand{\paperauthorcontributions}{%
  The author is solely responsible for this article.%
}

\title{\papertitle \\[0.35em]
\normalsize \papersubtitle}

\author{{\fontsize{8}{10}\selectfont\paperauthors}\\[0.9em]
{\scriptsize\paperaffiliations}}
\date{\vspace{1.5em}}

\begin{document}
\maketitle

\begin{abstract}
This paper proposes a formal multi-agent architecture for implementing enterprise AI in regulated insurance firms, integrating economic theory with institutional design. The framework synthesises three core theoretical perspectives: Arrow’s risk pooling theory to formalise risk transformation under uncertainty, Nash equilibrium to model strategic interactions between decision agents, and Principal-Agent theory to address incentive alignment under information asymmetry. The insurer is modelled as a constrained optimisation entity operating under solvency, legal, ESG, and operational boundaries, with specific focus on the regulatory contexts of Austria and Germany. The architecture decomposes the firm into multiple specialised agents—each representing distinct functional domains such as capital management, underwriting, claims processing, compliance, fraud detection, and client interaction. Human-in-the-loop agents are integrated through a tiered access control system, ensuring differentiated data visibility and decision influence based on user roles. An orchestrator agent supervises inter-agent coordination, enforcing regulatory admissibility and institutional coherence under frameworks such as Solvency II, the AI Act, and the Insurance Distribution Directive. Protocol integration is based on asynchronous execution and dual-layer communication infrastructures, specifically the Model Context Protocol (MCP) and Agent-to-Agent (A2A) messaging. This structure enables the systematic design of compliant, auditable multi-agent systems aligned with the institutional logic of financial firms in Austria and Germany.
\end{abstract}

\keywords{Multi-Agent Systems, Enterprise AI, Insurance Firms, Solvency II, AI Act, Regulated Environments, Constrained Optimisation, Principal-Agent Theory, Nash Equilibrium, Arrow’s Risk Pooling, Austria, Germany, Institutional Design, ESG Compliance, Regulatory Architecture, Model Context Protocol (MCP), Agent-to-Agent Protocol (A2A), AI Governance, Algorithmic Accountability, Financial Regulation
}

\bigskip

\section{Introduction}\label{sec:introduction}

In this paper, we develop a theoretical framework to formalise the core institutional functions of insurance companies operating in Austria and Germany. Our objective is to represent these firms as systems of constrained optimisation problems, suitable for implementation via a decentralised multi-agent architecture. We deliberately avoid task-level AI applications and instead focus on the structural transformation of insurance firms into formal agent-based systems governed by economic objectives and regulatory constraints. The analysis is restricted to profit-oriented private insurers, enabling the derivation of objective functions consistent with utility or surplus maximisation under bounded rationality. This work contributes to a foundation for applying advanced AI systems to insurance by grounding agent behaviour in firm theory rather than heuristic rule sets or isolated prediction models.

\subsection{Institutional Scope: Insurance Markets in Austria and Germany}\label{sec:introduction-scope}

Austria and Germany constitute two of the most developed and stringently regulated insurance markets in Europe. Both jurisdictions operate under the Solvency II framework, characterised by high market penetration, standardised supervisory practices, and formally codified capital adequacy regimes. The German market includes a significant share of mutuals and public-law insurers, whereas Austria exhibits stronger concentration among private joint-stock firms. Regulatory supervision is exercised by BaFin in Germany and the FMA in Austria, under harmonised EU directives. These institutional similarities—amplified by coercive and normative isomorphic pressures within the EU regulatory field—permit a joint theoretical treatment of profit-driven insurer behaviour \cite{DiMaggio1983}. Capital requirements and solvency constraints are explicitly quantified under Solvency II, making it possible to model insurance firms as surplus-maximising entities operating under regulatory risk boundaries \cite{Boonen2017}.

\subsection{Taxonomy of Insurance Firms}\label{sec:introduction-taxonomy}

Insurance firms in Austria and Germany can be categorised according to ownership structure, legal form, and regulatory mandate. Broadly, three types dominate the institutional landscape: (i) public-law insurers, such as statutory health insurance providers; (ii) mutual insurers, owned by their policyholders; and (iii) joint-stock companies, operating under shareholder control. While all entities are subject to Solvency II and supervised by national authorities, their internal objectives differ substantially. Public-law insurers are typically constrained by statutory duties and fixed benefit structures, with limited discretion over pricing or capital strategy. Mutuals operate under collective ownership, often prioritising member benefits over profit maximisation. In contrast, joint-stock insurers pursue surplus generation under market competition, subject to regulatory capital and risk constraints \cite{SchmeiserGruendl2002, Biener2016}. These differences affect the formalisation of objective functions and admissible decision sets. Given this divergence, the subsequent analysis focuses exclusively on profit-driven private insurers, for which the firm can be modelled as a utility-maximising entity operating under solvency and compliance constraints.

\subsection{Related Work}\label{sec:introduction-related}

The theoretical foundation for modelling insurer behaviour rests on three principal frameworks. First, Arrow’s treatment of risk-bearing institutions provides a canonical basis for interpreting insurance as a mechanism for intertemporal risk transfer and welfare improvement under uncertainty \cite{Arrow1963}. Second, the principal-agent framework is widely used to model asymmetries between insurers and policyholders, particularly in underwriting and claims settlement, where hidden information and hidden actions affect contract outcomes \cite{Rothschild1976, zweifel2012}. Third, the Nash equilibrium concept enables formal analysis of insurer interaction in competitive environments, such as premium setting, reinsurance strategy, or capital allocation, where strategic interdependence governs outcomes \cite{Dickson2004}. These foundations allow insurer behaviour to be framed as a set of constrained optimisation problems under uncertainty and regulation. This abstraction is essential for translating institutional logic into formal structures usable in multi-agent systems.

Beyond economic formalism, the internal structure and governance of insurance firms have been analysed through organisational theory. Principal-agent models extend to internal hierarchies, capturing incentive misalignments between shareholders, executives, and operational units \cite{Cummins2006}. Institutional isomorphism has been observed in the convergence of firm structures across markets with similar regulatory constraints, reinforcing the adoption of standardised roles and procedures \cite{DiMaggio1983}. This theoretical lens provides a basis for decomposing insurer operations into semi-autonomous subfunctions governed by compliance and incentive compatibility. Such decomposition aligns naturally with agent-based modelling approaches, in which decentralised decision entities operate under bounded rationality within a shared institutional environment.

Applications of AI in the insurance industry have focused primarily on operational subdomains such as fraud detection, customer profiling, and claims triage. Supervised learning techniques dominate, often relying on structured policyholder data or claim records \cite{Brockett2002}. More recently, natural language processing has been applied to extract and classify information from unstructured documents in claims workflows \cite{kolambe2023}. Industry-driven approaches increasingly promote intelligent automation platforms that combine machine learning with robotic process automation for end-to-end claims handling \cite{automationanywhere2021}. Such implementations remain task-specific and do not extend to the systemic modelling of insurer behaviour. Existing AI solutions typically optimise local performance metrics without incorporating institutional objectives or regulatory constraints, limiting their suitability for architecture-level integration.

\subsection{Contribution}\label{sec:introduction-contribution}

This paper contributes a theoretical framework that formalises the core institutional functions of profit-driven insurance firms as constrained optimisation problems grounded in economic theory. Building on established models of utility, risk, and contract theory, we decompose insurer operations into analytically distinct subfunctions aligned with firm-level objectives. These subfunctions are then mapped onto a multi-agent system architecture, in which autonomous agents operate under bounded rationality, institutional constraints, and shared utility goals.  Unlike task-oriented AI applications, the proposed structure enables systemic modelling of insurance firms as distributed decision systems, opening a path toward compliant, architecture-level AI implementation in regulated environments.

\section{Formal Modelling of Insurance Firm Functions}\label{sec:modelling}

The formal behaviour of an insurance company can be represented as an optimisation problem in which the firm seeks to maximise surplus or expected utility, subject to a set of internal and external constraints. The objective reflects the insurer’s economic role as an intermediary that transforms individual risks into collective stability through pooling and capital management. Constraints arise from capital requirements, risk boundaries, and regulatory compliance obligations, which delimit the admissible decision space. In regulated environments such as Austria and Germany, these constraints include solvency directives such as Solvency II, which formalise market-consistent valuation, capital adequacy, and risk-based supervision \cite{Starita2014,Rae2017}, consumer protection rules under the Insurance Distribution Directive (IDD), which aim to enhance transparency, improve advisory standards, and strengthen product governance in insurance distribution \cite{EIOPA2022,Werner2021}, and increasingly also algorithmic accountability derived from the AI Act, the GDPR, and ESG-related regulatory frameworks \cite{EuropeanCommission2021,SFDR2020}. This abstraction reduces the insurance firm to a constrained optimisation entity, forming the analytical basis for decomposing its internal functions into subcomponents represented by decision agents. As a baseline, the firm’s behaviour can be formalised as the following maximisation problem:
\begin{align}
\max_{\boldsymbol{a} \in \mathcal{A}} \quad & \mathbb{E}[U(\pi(\boldsymbol{a}))] \label{eq:firm-obj} \\
\text{s.t.} \quad
& \mathbb{E}[L(\boldsymbol{a})] \leq C \quad \text{(Capital adequacy constraint)} \label{eq:capital} \\
& \text{VaR}_{\alpha}(\pi(\boldsymbol{a})) \leq R \quad \text{(Risk exposure limit under Solvency II)} \label{eq:solvency} \\
& \boldsymbol{a} \in \mathcal{D}_{\text{legal}} \quad \text{(Regulatory admissibility: AI Act, GDPR, IDD)} \label{eq:legal} \\
& \boldsymbol{a} \in \mathcal{E}_{\text{ESG,E}} \quad \text{(Environmental screening under EU Taxonomy and SFDR)} \label{eq:esg-env} \\
& \boldsymbol{a} \in \mathcal{G}_{\text{ESG,S\&G}} \quad \text{(Social and governance constraints)} \label{eq:esg-sg} \\
& \boldsymbol{a} \in \mathcal{O}_{\text{operational}} \quad \text{(Operational feasibility)} \label{eq:operational}
\end{align}

In this formulation, $\boldsymbol{a} \in \mathcal{A}$ denotes the vector of firm-level decisions, including pricing, underwriting, claims handling, investment allocation, and governance policies. The function $\pi(\boldsymbol{a})$ represents the surplus or profit generated by these decisions, while $U(\cdot)$ is the firm's utility function, potentially reflecting risk-neutral or risk-averse preferences.

The capital adequacy constraint $\mathbb{E}[L(\boldsymbol{a})] \leq C$ ensures that expected liabilities do not exceed available capital $C$. Risk exposure is formalised via a Value-at-Risk condition at confidence level $\alpha$, such that $\text{VaR}_{\alpha}(\pi(\boldsymbol{a})) \leq R$, where $R$ is the firm's Solvency II risk threshold. The admissible decision space is further bounded by legal and institutional requirements. The legal constraint $\boldsymbol{a} \in \mathcal{D}_{\text{legal}}$ enforces compliance with supervisory regimes such as the AI Act~\cite{EuropeanCommission2021}, the General Data Protection Regulation (GDPR), and the Insurance Distribution Directive (IDD)~\cite{EIOPA2022}. ESG admissibility is introduced via two distinct constraint sets: $\boldsymbol{a} \in \mathcal{E}_{\text{ESG,E}}$ denotes environmental eligibility under instruments such as the EU Taxonomy Regulation (EU) 2020/852~\cite{EuropeanParliament2020} and the Sustainable Finance Disclosure Regulation (SFDR)~\cite{SFDR2019}, including measurable thresholds like carbon intensity per revenue unit or share of taxonomy-aligned investments. $\boldsymbol{a} \in \mathcal{G}_{\text{ESG,S\&G}}$ ensures that actions meet minimum social and governance criteria, including metrics such as board diversity ratios, executive pay dispersion, transparency of grievance mechanisms, and policyholder participation in governance. $\boldsymbol{a} \in \mathcal{O}_{\text{operational}}$ guarantees that all selected actions remain within the firm’s technical and procedural capabilities.

This formulation expresses the insurer’s decision logic as a constrained optimisation problem: maximising expected utility while navigating capital limits, solvency thresholds, regulatory boundaries, and ESG accountability. It reflects the institutional reality that insurance firms are not profit-maximising entities in a vacuum, but regulated intermediaries embedded within a system of legal, financial, and sustainability norms. The constraint structure provides a formal foundation for balancing these competing obligations within an integrated decision architecture. The constrained optimisation problem can be reformulated using a Lagrangian representation, which incorporates the firm’s decision space and associated constraints into a single augmented objective:
\begin{align}
\mathcal{L}(\boldsymbol{a}, \boldsymbol{\lambda}) = \mathbb{E}[U(\pi(\boldsymbol{a}))] - \lambda_1 \left( \mathbb{E}[L(\boldsymbol{a})] - C \right) - \lambda_2 \left( \text{VaR}_{\alpha}(\pi(\boldsymbol{a})) - R \right) - \sum_{i=3}^{6} \lambda_i \cdot \mathbb{I}_{\text{viol}}^{(i)}(\boldsymbol{a})
\end{align}

In this formulation, $\boldsymbol{\lambda} = (\lambda_1, \lambda_2, \ldots, \lambda_6)$ denotes the vector of non-negative Lagrange multipliers associated with the respective constraints. The indicator functions $\mathbb{I}_{\text{viol}}^{(i)}(\boldsymbol{a})$ take the value 1 when the corresponding constraint is violated, and 0 otherwise. This representation allows each binding constraint to be interpreted as an implicit cost component: a marginal penalty or trade-off that reduces the firm’s attainable utility. The multipliers quantify the shadow price of each constraint — i.e. how much expected utility would improve if that constraint were marginally relaxed.

Put differently, this expression formalises the reality that insurers are not just maximising profit; they are doing so under legal, financial, and institutional boundaries, each of which carries a hidden opportunity cost. The formulation makes those trade-offs explicit and measurable.

This prepares the ground for decomposing the global optimisation problem into analytically tractable subcomponents, each corresponding to a distinct functional domain within the firm. These subcomponents will later be mapped to autonomous decision agents, coordinated within a multi-agent system aligned with the firm’s overall objective.

The formal constraint structure outlined above provides a general decision-theoretic foundation. To refine it, we now examine how different theoretical frameworks contribute distinct modelling perspectives to insurer behaviour. We begin with Arrow’s treatment of risk-bearing institutions. Arrow’s analytical framework interprets insurance as a mechanism for transferring individual uncertainty into collective stability through risk pooling under conditions of incomplete information and risk aversion \cite{Arrow1963}.

Within this view, the firm functions as a utility-transforming intermediary, accepting idiosyncratic risks from clients and aggregating them into diversified portfolios whose outcomes are more predictable at the collective level. Risk-averse agents maximise the expected utility of their final wealth, implying that the insurance contract must improve the expected utility of the insured while maintaining the insurer’s solvency. The firm's objective thus incorporates a concave utility function, and optimal contract design becomes a question of balancing marginal utility across risk classes subject to regulatory and capital constraints. This foundation legitimises the use of expected utility in the global model and provides a formal link between micro-level risk aversion and firm-level surplus transformation.

Following the expected utility framework established by \textcite{Arrow1963}, if $W$ denotes the insurer’s terminal wealth and $X$ the aggregate uncertain claim distribution, the firm solves:
\begin{align}
\max \quad & \mathbb{E}[U(W - X)] \label{eq:arrow}
\end{align}

where $U(\cdot)$ is a strictly concave utility function reflecting risk aversion, and $X$ represents the net realised liabilities from pooled insured risks. This structure underpins the expected utility formulation applied in the global optimisation model.

A second theoretical perspective relevant to insurer modelling derives from principal–agent theory, which formalises the implications of asymmetric information between contracting parties. In insurance, such asymmetries are structural: the policyholder holds private information about risk type (adverse selection) and actions taken post-contract (moral hazard) \cite{Rothschild1976}. This leads to inefficiencies in underwriting, pricing, and claims settlement, where the insurer must design mechanisms to extract truthful signals or induce appropriate behaviour.

Within the firm itself, principal–agent problems arise in governance structures, where shareholders (principals) must align the actions of executives and operational units (agents) under limited observability and incentive misalignment \cite{Cummins2006}. In the modelling context, principal–agent theory justifies the use of informational constraints in the firm’s decision space. These constraints restrict the feasible set of actions not only by regulation and capital, but by the need to maintain incentive compatibility.

The global optimisation problem is thus shaped by contracts and internal mechanisms that ensure delegated decisions align with firm objectives despite decentralised knowledge and interests. Following the principal-agent framework formalised by \textcite{Rothschild1976}, an incentive-compatible mechanism can be expressed as:
\begin{align}
a^* \in \arg\max_{a \in \mathcal{A}} \ \mathbb{E}[u(a, s(a)) \mid \theta] \label{eq:ic}
\end{align}

where $a^*$ is the equilibrium action, $u(\cdot)$ is the agent’s utility, and $\theta$ is the agent’s private type. This structure embeds information asymmetry into the admissible decision structure of the firm.   Similarly, Nash equilibrium provides the formal language to describe strategic interactions among insurers in competitive markets, as outlined by \textcite{Dickson2004}. Each firm chooses a strategy that maximises its objective given the strategies of others.

This strategic context introduces equilibrium constraints into the firm’s optimisation problem, particularly in areas where regulatory frameworks permit competition under capital requirements and fair disclosure. Premium rates, for example, are not solely determined by loss expectations, but also by competitive pressures and reactions. Similarly, capital allocation and reinsurance retention levels are shaped by the structure of rival firms’ positions. In this setting, the insurer’s feasible set is partially endogenous: constrained not only by internal and external norms, but by strategic stability.

Following the Nash equilibrium framework as applied to insurance markets by \textcite{Dickson2004}, the equilibrium condition can be formalised as:
\begin{align}
\boldsymbol{a}_i^* \in \arg\max_{\boldsymbol{a}_i \in \mathcal{A}_i} \ \mathbb{E}[U_i(\pi_i(\boldsymbol{a}_i, \boldsymbol{a}_{-i}))] \label{eq:nash}
\end{align}

where each firm maximises its own expected utility conditional on the strategies of its competitors. This condition defines a Nash equilibrium in which no insurer has an incentive to deviate unilaterally from its chosen action.

The three modelling perspectives outlined above—Arrow’s utility-based risk pooling \parencite{Arrow1963}, principal–agent theory under asymmetric information \parencite{Rothschild1976}, and Nash equilibrium in strategic competition \parencite{Dickson2004}—each contribute distinct structural constraints and behavioural mechanisms to the insurer’s decision space. Together, they represent a layered interpretation of insurance firm behaviour: risk transformation, incentive design, and interdependent strategy selection.

To capture this multidimensional structure, we now introduce a unified formulation in which the firm is modelled as a distributed decision system composed of interdependent subfunctions. This approach allows each theoretical lens to be encoded as a constraint or objective component in a global architecture that reflects the institutional and economic realities of regulated insurance markets. It also provides a formal foundation for decomposing the firm into agent-level components in later sections. We define the unified model as follows:
\begin{align}
\max_{\boldsymbol{a}_1, \dots, \boldsymbol{a}_n} \quad & \sum_{i=1}^n \mathbb{E}[U_i(\pi_i(\boldsymbol{a}_i, \boldsymbol{a}_{-i}))] \label{eq:unified-obj} \\
\text{s.t.} \quad
& \mathbb{E}[L_i(\boldsymbol{a}_i)] \leq C_i \quad \forall i \quad \text{(Capital adequacy)} \label{eq:unified-capital} \\
& \text{IC}_i(\boldsymbol{a}_i, \theta_i) \quad \text{holds} \quad \forall i \quad \text{(Incentive compatibility)} \label{eq:unified-ic} \\
& \boldsymbol{a}_i \in \mathcal{B}_i(\boldsymbol{a}_{-i}) \quad \forall i \quad \text{(Best response: strategic consistency)} \label{eq:unified-bestresponse} \\
& \boldsymbol{a}_i \in \mathcal{D}_i \cap \mathcal{E}_i \cap \mathcal{G}_i \cap \mathcal{O}_i \quad \forall i \quad \text{(Legal, ESG, operational admissibility)} \label{eq:unified-constraints}
\end{align}

Here, the firm is composed of $n$ subfunctions or decision units, each denoted by index $i$. Each $\boldsymbol{a}_i$ represents a vector of decisions associated with a particular function (e.g. pricing, underwriting, claims). The function $\pi_i(\boldsymbol{a}_i, \boldsymbol{a}_{-i})$ denotes the financial contribution of function $i$, which may depend on its own decisions as well as those of other units (interdependencies). $U_i(\cdot)$ is the local utility function, potentially reflecting risk preferences or performance targets. Constraint \eqref{eq:unified-capital} ensures local capital adequacy; \eqref{eq:unified-ic} imposes incentive compatibility under private information; \eqref{eq:unified-bestresponse} enforces strategic consistency in the presence of inter-agent competition or coordination; and \eqref{eq:unified-constraints} collects external admissibility constraints from regulatory, environmental, governance, and operational domains.

This structure treats an insurance company as a system made up of smaller expert units. Each unit has its own job and goals, but they all work together within a shared set of rules and limits. Some rules come from regulators (like Solvency II or the AI Act), others from the market or the firm's own governance. The model shows how each part makes decisions, while still aligning with the company's overall direction. This formalisation sets the stage for implementing a multi-agent system that reflects how real insurers actually function under legal, financial, and strategic constraints.

\section{Functional Decomposition and Agent Architecture}\label{sec:agent-structure}

Given the regulatory, confidentiality, and institutional constraints of the insurance sector, we argue that enterprise AI systems must be implemented as proprietary infrastructures rather than built on open, general-purpose platforms. This requirement arises from the intersection of data protection obligations (GDPR), auditability under regulatory supervision (IDD, AI Act), and the need for institution-specific objective functions. Within this context, we propose a decentralised multi-agent architecture as the appropriate system design, in which autonomous decision agents operate under a global regulatory and strategic logic enforced by an overseeing orchestrator agent.

The complexity of modern insurance firms necessitates a modular architecture, where distinct functions operate as autonomous, goal-directed subunits. This decomposition enhances both analytical tractability and system design, particularly when formalising insurer behaviour as a multi-agent system. Each agent is assigned a specific functional role, operating under local decision rules bounded by firm-level constraints. While agents such as underwriting, capital management, and claims settlement act semi-independently, they are coordinated through a global optimisation logic that enforces regulatory admissibility, risk exposure boundaries, and capital adequacy. Human-in-the-loop agents are integrated through a tiered access system, structuring data visibility and decision authority based on user roles, ensuring both regulatory compliance and operational alignment.

In addition to technical operations, the firm must satisfy ESG-related obligations, particularly in the social and governance (S\&G) domains. This includes traceable decision accountability, inclusive stakeholder access, and transparent reporting structures. To this end, some agents may not optimise financial flows directly, but rather enforce compliance, track governance metrics, or interface with external actors. Large language models (LLMs) provide a natural interface layer for employee- and client-facing communication, enabling agents to interact with users in transparent, interpretable terms. These LLM-enabled agents can also serve stakeholders such as shareholders or regulators by offering on-demand conversational access to policy logic, performance summaries, and risk assessments. In this setting, natural language becomes both a user interface and a compliance instrument, supporting the broader governance objectives of the firm.

As a profit-oriented institution, the insurance firm seeks to maximise financial surplus subject to internal and regulatory constraints.

\subsection*{Capital Management Agent}

The capital management agent is responsible for maintaining solvency while supporting business growth. Its function is to allocate capital across risk-bearing units, ensure compliance with solvency requirements (e.g. Solvency II), and buffer against adverse shocks. The agent does not generate profit directly but enables all other agents to operate within admissible financial boundaries. Its decision problem balances reserve levels, liquidity availability, and capital costs. Formally, the capital agent solves:
\begin{align}
\max_{k \in \mathcal{K}} \quad & -\gamma \cdot \text{Cost}(k) \label{eq:capital-obj} \\
\text{s.t.} \quad
& \mathbb{P}(X > k) \leq 1 - \alpha \quad \text{(Solvency constraint)} \label{eq:capital-solvency} \\
& k \geq \text{Reserve}_{\text{min}} \quad \text{(Regulatory reserve floor)} \label{eq:capital-floor} \\
& k \leq \text{AvailableCapital} \quad \text{(Capital availability)} \label{eq:capital-available}
\end{align}

Here, $k$ denotes the capital buffer allocated by the agent, and $\text{Cost}(k)$ is a convex function representing the opportunity cost or regulatory friction of holding capital. $\gamma$ is a weighting coefficient that captures the trade-off between safety and efficiency. $X$ is the random variable denoting aggregate liability. The solvency constraint \eqref{eq:capital-solvency} ensures that the probability of insolvency remains below a predefined risk threshold $\alpha$ (e.g. 99.5\% for Solvency II). Constraint \eqref{eq:capital-floor} imposes a reserve floor, and \eqref{eq:capital-available} limits decisions to available financial resources.

In plain terms, this agent decides how much capital the insurer should hold in reserve so that it can stay solvent during bad years, while also not tying up too much money that could otherwise be used for business. It plays a safety role in the system, making sure the company doesn't fall below legal or supervisory thresholds, while maintaining sufficient financial flexibility to support underwriting and operational activities.

\subsection*{Underwriting Agent}

The underwriting agent is tasked with selecting, pricing, and classifying risks submitted to the insurer. Its goal is to accept profitable risks, reject adverse ones, and allocate fair premiums in line with expected losses and capital requirements. This agent faces information asymmetries due to incomplete or biased disclosures by applicants, and must therefore rely on observable indicators or probabilistic models. It interacts closely with the pricing function, fraud detection, and capital allocation. Formally, the underwriting agent solves:
\begin{align}
\max_{\boldsymbol{u} \in \mathcal{U}} \quad & \sum_{j=1}^m \left( p_j - \mathbb{E}[l_j \mid \boldsymbol{x}_j] \right) \cdot \delta_j \label{eq:underwriting-obj} \\
\text{s.t.} \quad
& \delta_j \in \{0,1\} \quad \forall j \quad \text{(Acceptance decision)} \label{eq:underwriting-accept} \\
& \mathbb{E}[l_j \mid \boldsymbol{x}_j] \leq \tau \quad \text{(Expected loss threshold)} \label{eq:underwriting-threshold} \\
& \sum_{j=1}^m \delta_j \cdot \text{VaR}_\alpha(l_j) \leq k \quad \text{(Capital exposure constraint)} \label{eq:underwriting-capital}
\end{align}

In this formulation, $j$ indexes the incoming applications, $p_j$ is the proposed premium, $l_j$ is the stochastic loss for applicant $j$, and $\boldsymbol{x}_j$ is the observable feature vector. $\delta_j$ is a binary variable that indicates whether the policy is accepted. The agent maximises expected underwriting profit across accepted applications while ensuring that no individual risk exceeds an expected loss threshold $\tau$ and that aggregate accepted risks do not violate the firm’s capital limit $k$ under VaR-based solvency requirements.

Put simply, this agent decides which applications to accept and at what premium. It tries to find a balance: avoiding customers likely to generate large losses, pricing correctly for those accepted, and keeping the portfolio within the firm’s risk capacity. It acts as the firm's first line of financial defence.

\subsection*{Claims Handling Agent}

The claims handling agent is responsible for assessing, verifying, and settling claims made by policyholders. Its role is to ensure fair and timely payment of legitimate claims while preventing overcompensation, delay-induced escalation, or exposure to fraudulent activity. It operates under uncertainty due to incomplete documentation and the stochastic nature of claim sizes and timing. It coordinates with the fraud detection agent, legal interface, and capital management unit. Formally, the claims agent solves:
\begin{align}
\min_{\boldsymbol{c} \in \mathcal{C}} \quad & \sum_{i=1}^n \mathbb{E}[P_i(\theta_i)] + \lambda \cdot \text{Delay}_i(\theta_i) \label{eq:claims-obj} \\
\text{s.t.} \quad
& P_i(\theta_i) \geq \underline{C}_i \quad \forall i \quad \text{(Minimum contractual payout)} \label{eq:claims-min} \\
& P_i(\theta_i) \leq \overline{C}_i \quad \forall i \quad \text{(Policy coverage limit)} \label{eq:claims-max} \\
& \text{FraudScore}_i(\theta_i) \leq \phi \quad \forall i \quad \text{(Fraud detection screen)} \label{eq:claims-fraud}
\end{align}

Here, $P_i(\theta_i)$ is the payout for claim $i$ as a function of latent claim type $\theta_i$, which is uncertain at the time of processing. $\underline{C}_i$ and $\overline{C}_i$ denote the policy’s minimum and maximum payout obligations. $\text{Delay}_i(\theta_i)$ represents cost associated with processing time (which may include penalties or legal escalation). The fraud score is an externally or internally computed index bounded by $\phi$, above which claims are flagged for additional screening.

In simple terms, this agent manages the decision of how much to pay out and when. It tries to fulfil the firm's promises to policyholders but must be careful not to pay more than contractually required—or too quickly if the claim appears suspicious. Its job is about fairness, vigilance, and legal correctness under time pressure.

\subsection*{Fraud Detection Agent}

The fraud detection agent operates in parallel with the claims and underwriting processes, monitoring for anomalous patterns or inconsistent declarations that may indicate intentional misrepresentation. Its purpose is to identify and flag high-risk cases for further review, thereby reducing exposure to internal and external manipulation. It leverages probabilistic models, anomaly detection algorithms, or learned patterns to assess the integrity of incoming data. This agent is preventive in nature and does not directly affect financial outcomes but modifies the decision space of other agents. Formally, the fraud agent solves:
\begin{align}
\max_{\boldsymbol{f} \in \mathcal{F}} \quad & \sum_{i=1}^n \mathbb{E}[\text{TP}_i(\theta_i)] - \mu \cdot \mathbb{E}[\text{FP}_i(\theta_i)] \label{eq:fraud-obj} \\
\text{s.t.} \quad
& \text{Score}_i(\theta_i) = \mathcal{M}(\boldsymbol{x}_i) \quad \forall i \quad \text{(Fraud model output)} \label{eq:fraud-score} \\
& \text{Flag}_i = \mathbb{I}[\text{Score}_i(\theta_i) > \phi] \quad \forall i \quad \text{(Threshold decision)} \label{eq:fraud-flag}
\end{align}

In this formulation, $\text{TP}_i$ and $\text{FP}_i$ represent the expected true and false positives for claim or policy $i$ under the agent’s detection regime. $\mu$ is a penalty parameter on false positives to discourage excessive filtering. $\mathcal{M}(\cdot)$ is a predictive model mapping observable features $\boldsymbol{x}_i$ to a fraud score. Claims or applications are flagged if their score exceeds a risk threshold $\phi$.

In simple terms, this agent watches for red flags. It uses data to guess which claims or applications might be fake or manipulated. If the score is too high, the case gets flagged for deeper review. It helps protect the insurer from bad actors without blocking genuine customers.

\subsection*{Compliance and Legal Agent}

The compliance and legal agent ensures that all internal decisions and external product features adhere to applicable regulations and governance norms. It enforces admissibility under frameworks such as Solvency II, the Insurance Distribution Directive (IDD), the General Data Protection Regulation (GDPR), the AI Act, and ESG-related disclosure rules. This agent does not optimise profit directly, but constrains the action space of all other agents by validating whether decisions are legally and ethically permissible.
Formally, the agent implements a filtering operator:
\begin{align}
\forall i: \quad \boldsymbol{a}_i^{\text{final}} =
\begin{cases}
\boldsymbol{a}_i & \text{if } \boldsymbol{a}_i \in \mathcal{D}_{\text{legal}} \cap \mathcal{E}_{\text{ESG,E}} \cap \mathcal{G}_{\text{ESG,S\&G}} \\
\emptyset & \text{otherwise}
\end{cases} \label{eq:compliance-filter}
\end{align}

Here, $\boldsymbol{a}_i$ denotes the proposed action from agent $i$, and $\boldsymbol{a}_i^{\text{final}}$ is the admissible action after legal and ethical screening. The operator enforces exclusion if any rule—legal, environmental, social, or governance—is violated. ESG-S\&G includes procedural fairness, transparency, stakeholder inclusion, and traceability of decisions.

In simple terms, this agent acts as the internal regulator. It checks whether actions suggested by other agents—such as new products, capital decisions, or underwriting rules—are legally allowed and ethically sound. If not, the action is blocked. It ensures the company plays by the rules and maintains reputational integrity.

\subsection*{Employee Interface Agent}

The employee interface agent serves as a controlled gateway between personnel and the AI-based decision architecture of the insurance system. Its objective is to maximise the individual work utility of each employee by providing context-relevant access to agent outputs—such as those from underwriting, claims, or compliance—while enforcing strict internal control over data visibility, traceability, and policy adherence. Each employee $e \in \mathcal{E}$ operates within an organisational role $r \in \mathcal{R}$, which defines their permitted query set and data access tier. The agent facilitates system interaction through a filtered interface, while ensuring compliance with data protection obligations (e.g. GDPR), internal governance rules, and audit requirements. Formally, the agent solves the following constrained optimisation problem for each employee:
\begin{align}
\max_{\boldsymbol{q}_e \in \mathcal{Q}_r} \quad & \mathbb{E}\left[U_e\left(\mathcal{L}(\boldsymbol{q}_e, \boldsymbol{a}_{\text{context}}^r)\right)\right] \label{eq:employee-obj} \\
\text{s.t.} \quad
& r \in \mathcal{R}_{\text{permitted}}(e) \quad \text{(Role-based access)} \label{eq:employee-role} \\
& \boldsymbol{a}_{\text{context}}^r \subseteq \boldsymbol{A}_{\text{global}} \quad \text{(Tier-constrained context)} \label{eq:employee-context} \\
& \text{log}(\boldsymbol{q}_e, t, e) \in \mathcal{L}_{\text{audit}} \quad \text{(Supervisory auditability)} \label{eq:employee-audit}
\end{align}

Here, $\boldsymbol{q}_e$ denotes a query submitted by employee $e$, and $\boldsymbol{a}_{\text{context}}^r$ is the filtered view of the agent system state accessible under role $r$. The global state $\boldsymbol{A}_{\text{global}}$ comprises all structured outputs from other agents. The function $\mathcal{L}(\cdot)$ is the LLM interface that maps decision-relevant data to human-readable outputs. The utility function $U_e(\cdot)$ reflects the employee’s expected task efficiency or decision quality from receiving this response. Constraint~\eqref{eq:employee-role} ensures that the role is authorised; \eqref{eq:employee-context} enforces tier-based data isolation; and \eqref{eq:employee-audit} requires that every query is logged, timestamped, and auditable.

In simpler terms, this setup ensures that each employee gets exactly the information they need to do their job—no more, no less. Their role in the company defines what they are allowed to see. All their requests are checked, filtered, and recorded so that sensitive data stays protected and everything remains transparent and traceable. To operationalise the tiered system, the agent applies a conditional access mapping:
\begin{align}
\text{Response}_e =
\begin{cases}
\mathcal{L}(\boldsymbol{q}_e, \boldsymbol{a}_{\text{context}}^r) & \text{if } r \in \mathcal{R}_{\text{permitted}}(e) \\
\emptyset & \text{otherwise}
\end{cases}
\label{eq:employee-tier}
\end{align}

This structure ensures that only admissible query–role combinations produce responses. For example, a call centre employee might access only claim-level summaries or policy status updates, whereas compliance officers can view audit logs, and executives are granted high-level system metrics and inter-agent coordination summaries. Role hierarchies are enforced dynamically through the internal access control engine, which references $\mathcal{R}_{\text{permitted}}(e)$ and monitors all activity under $\mathcal{L}_{\text{audit}}$. In functional terms, the employee interface agent enables precise, lawful, and productive human-AI interaction across the organisational hierarchy.

Put simply, the system checks who is asking and only gives an answer if the person’s role allows it. A junior employee might see only the basics needed for their task, while senior staff get broader insights. Everything is filtered automatically so that access always matches the person's position and responsibility in the company.

\subsection*{Stakeholder Interface Agent}

The stakeholder interface agent manages external access to the AI system by institutional and individual actors such as policyholders, regulators, auditors, shareholders, and business partners. Each stakeholder class is assigned a predefined visibility tier based on its contractual position, regulatory entitlement, or information rights. The agent ensures that external queries are resolved within lawful, contract-compliant, and context-specific boundaries, reflecting obligations under ESG-S\&G standards—particularly explainability, procedural fairness, and inclusive governance. The objective of the stakeholder interface agent is to maximise the relevance and interpretability of information provided to authorised external users, without breaching confidentiality or exceeding regulatory limits. Each stakeholder $s \in \mathcal{S}$ is mapped to a visibility tier, which governs their admissible query set and contextual scope. Formally, the agent solves:
\begin{align}
\max_{\boldsymbol{q}_s \in \mathcal{Q}_s} \quad & \mathbb{E}\left[U_s\left(\mathcal{L}(\boldsymbol{q}_s, \boldsymbol{a}_{\text{context}}^s)\right)\right] \label{eq:stakeholder-obj} \\
\text{s.t.} \quad
& s \in \mathcal{S}_{\text{authorised}} \quad \text{(Stakeholder authentication)} \label{eq:stakeholder-auth} \\
& \boldsymbol{a}_{\text{context}}^s \subseteq \boldsymbol{A}_{\text{global}} \quad \text{(Tiered data scope)} \label{eq:stakeholder-context} \\
& \text{log}(\boldsymbol{q}_s, t, s) \in \mathcal{L}_{\text{audit}} \quad \text{(Audit trace)} \label{eq:stakeholder-audit}
\end{align}

Here, $\boldsymbol{q}_s$ is a structured query submitted by stakeholder $s$, and $\mathcal{Q}_s$ is the set of queries permitted at that stakeholder’s access level. $\boldsymbol{a}_{\text{context}}^s$ is the filtered agent context available to that tier, constrained by $\boldsymbol{A}_{\text{global}}$, the complete internal state. The function $\mathcal{L}(\cdot)$ maps internal decisions into legally intelligible responses, adapted to jurisdiction, stakeholder contract, and purpose. The expected utility function $U_s(\cdot)$ captures the relevance and decision-usefulness of the returned information. Constraint~\eqref{eq:stakeholder-auth} enforces access authentication; \eqref{eq:stakeholder-context} restricts data exposure; and \eqref{eq:stakeholder-audit} ensures full supervisory traceability.

In simpler terms, this agent answers stakeholder questions using only the data they are legally allowed to see. It ensures that every external user—whether a customer, regulator, or investor—receives information that is relevant to their role, without exposing internal logic or private data. All interactions are tracked and audited to ensure lawful and accountable communication.

Stakeholder queries are processed under a conditional access mapping, where each stakeholder is assigned to a predefined access tier that determines which parts of the system they are permitted to view. This tiered structure reflects legal rights, contractual roles, and regulatory status, ensuring that each query is resolved only within the boundaries of the stakeholder’s assigned level.
\begin{align}
\text{Response}_s =
\begin{cases}
\mathcal{L}(\boldsymbol{q}_s, \boldsymbol{a}_{\text{context}}^s) & \text{if } s \in \mathcal{S}_{\text{authorised}} \\
\emptyset & \text{otherwise}
\end{cases}
\label{eq:stakeholder-tier}
\end{align}

This logic guarantees that each authorised stakeholder receives precisely the subset of information necessary for their role. A policyholder can access personal contracts, coverage parameters, and claim status. A regulator is entitled to solvency compliance data, ESG indicators, and audit trails. A shareholder may view profitability breakdowns, capital allocation, and long-run portfolio structure. Smart contract enforcement and role registries dynamically govern access rights under $\mathcal{S}_{\text{authorised}}$.

Put simply, this rule acts like a smart filter. Only stakeholders with proper authorisation receive tailored answers to their questions. Everyone else gets nothing. This prevents unauthorised access and ensures that each user sees only what they are meant to—no more, no less.

\subsection*{Client Service Agent}

The client service agent governs interactive access for existing policyholders, enabling personalised communication regarding policy coverage, claim progression, premium adjustments, renewals, and contractual terms. Its objective is to maximise customer clarity and decision-readiness while ensuring that all disclosures remain compliant with legal, contractual, and data protection boundaries. Each client $c \in \mathcal{C}$ is associated with a current contract, a policy data record $\boldsymbol{p}_c$, and a visibility tier based on policy status and jurisdiction. Formally, the client service agent solves:
\begin{align}
\max_{\boldsymbol{q}_c \in \mathcal{Q}_c} \quad & \mathbb{E}\left[U_c\left(\mathcal{L}(\boldsymbol{q}_c, \boldsymbol{p}_c, \boldsymbol{a}_{\text{status}})\right)\right] \label{eq:client-service-obj} \\
\text{s.t.} \quad
& c \in \mathcal{C}_{\text{active}} \quad \text{(Contract status)} \label{eq:client-active} \\
& \boldsymbol{p}_c \in \mathcal{P}_{\text{entitled}}(c) \quad \text{(Policy-bound visibility)} \label{eq:client-policy} \\
& \text{log}(\boldsymbol{q}_c, t, c) \in \mathcal{L}_{\text{audit}} \quad \text{(Interaction traceability)} \label{eq:client-audit}
\end{align}

Here, $\boldsymbol{q}_c$ is a query submitted by client $c$, and $\mathcal{Q}_c$ is the set of permitted queries under their active contract. The variable $\boldsymbol{p}_c$ represents their personalised policy attributes (e.g. coverage limits, deductibles, renewal terms), while $\boldsymbol{a}_{\text{status}}$ includes dynamic operational information such as claim updates or payment schedules. The LLM interface $\mathcal{L}(\cdot)$ converts structured internal data into accessible, legally compliant responses. The utility function $U_c(\cdot)$ reflects perceived information value, clarity, and support for action by the client. Constraints \eqref{eq:client-active} and \eqref{eq:client-policy} limit visibility to entitled content only, while \eqref{eq:client-audit} ensures that all exchanges are auditable.

In simpler terms, the system checks whether a person is a valid, active customer and then tailors the response based on their contract. It helps them understand their rights, monitor their claims, and make informed choices—while keeping a clear record of every request and answer. To implement this logic, the agent applies a conditional access mapping:
\begin{align}
\text{Response}_{c} =
\begin{cases}
\mathcal{L}(\boldsymbol{q}_c, \boldsymbol{p}_c, \boldsymbol{a}_{\text{status}}) & \text{if } c \in \mathcal{C}_{\text{active}} \\
\emptyset & \text{otherwise}
\end{cases}
\label{eq:client-service}
\end{align}

This structure ensures that only currently entitled policyholders receive access to personalised, contract-linked information. The visibility scope depends on the client’s policy type, current status (e.g. in-claim, pending renewal), and applicable legal protections. For example, a health insurance client in an open claim may query reimbursement timelines, while a property policyholder near renewal can request premium comparisons. The filtering logic reflects a tiered access model tied to dynamic contract context and compliance policy.

Put plainly, this agent acts like a personalised digital advisor: it checks if someone is still an active client and then answers questions based only on what they are allowed to see. It keeps things clear, relevant, and legal—without ever showing what doesn’t belong to them.

\subsection*{Client Acquisition Agent}

The client acquisition agent is responsible for engaging potential customers, mapping product offerings to prospect profiles, and pre-filtering applications for underwriting relevance. It operates upstream of the underwriting agent and integrates with marketing, pricing, and legal modules. Its role is to generate qualified leads, screen suitability, and deliver regulatory pre-contractual information in accessible language. The agent is constrained by marketing compliance rules and fairness requirements under anti-discrimination directives.

Formally, the agent implements a pre-qualification mapping:
\begin{align}
\boldsymbol{p}_{\text{suggested}} = \mathcal{M}_{\text{map}}(\boldsymbol{x}_{\text{prospect}}) \quad \text{if } \boldsymbol{x}_{\text{prospect}} \in \mathcal{X}_{\text{admissible}} \label{eq:client-acquisition}
\end{align}

Here, $\boldsymbol{x}_{\text{prospect}}$ contains declared or inferred features of a potential customer (e.g. age, location, coverage needs), and $\mathcal{M}_{\text{map}}(\cdot)$ maps these features to a product recommendation $\boldsymbol{p}_{\text{suggested}}$, subject to legal filters $\mathcal{X}_{\text{admissible}}$ (e.g. no discriminatory profiling, marketing restrictions). The agent also exposes LLM interfaces for interactive product comparison, consent dialogue, and onboarding FAQs.

In effect, this agent serves as the public face of the firm’s system, converting prospects into compliant applications. It educates, filters, and prepares the ground for underwriting by delivering structured pre-application guidance, all while ensuring that acquisition processes are transparent, equitable, and regulatorily compliant.

\subsection*{System Orchestrator Agent}

The system orchestrator agent serves as the supervisory layer responsible for ensuring global consistency, strategic alignment, and institutional coherence across all subordinate agents. It oversees multi-agent coordination, verifying that local decisions—while independently optimised—collectively satisfy the legal, financial, and operational constraints of the organisation. The orchestrator does not perform direct optimisation; instead, it acts as a global validator of admissibility, equilibrium, and inter-agent alignment. Formally, the orchestrator monitors whether the system-wide agent state $\{\boldsymbol{a}_1, \dots, \boldsymbol{a}_n\}$ respects both the global objective structure and admissibility envelope:
\begin{align}
\{\boldsymbol{a}_i\}_{i=1}^n \in \arg\max_{\boldsymbol{a}_1, \dots, \boldsymbol{a}_n} \ \sum_i \mathbb{E}[U_i(\pi_i(\boldsymbol{a}_i, \boldsymbol{a}_{-i}))] \quad \text{s.t. global feasibility and coordination constraints} \label{eq:orchestrator}
\end{align}

Its role is to verify system-wide admissibility and consistency and to authorise or reject execution based on validation:
\begin{align}
\text{Approved} =
\begin{cases}
1 & \text{if all admissibility, consistency, and coordination rules are met} \\
0 & \text{otherwise}
\end{cases}
\label{eq:orchestrator-flag}
\end{align}

The orchestrator integrates several supervisory mechanisms to enforce institutional coherence. First, it performs constraint propagation across agent boundaries, ensuring that local admissibility conditions do not conflict when aggregated at the system level. Second, it verifies role-permission consistency across agents, maintaining uniformity in how organisational roles are mapped to actions and data access within different subsystems. Third, it monitors for cross-agent conflicts—such as contradictory objectives, redundant actions, or mutual constraint violations—and triggers resolution logic when necessary. Finally, the orchestrator aligns internal agent decisions with external supervisory expectations and the firm’s strategic objectives, ensuring that the AI system acts in accordance with institutional goals and regulatory mandates.

It may override, veto, or delay decisions that would introduce institutional incoherence, constraint violations, or regulatory breaches. While the legal/compliance agent enforces rule-level admissibility for individual actions, the orchestrator governs **system-level integration** and strategic consistency. LLM modules attached to this agent may generate system summaries, policy explanations, or audit statements for institutional stakeholders (e.g. board, regulators).

In simpler terms, this agent makes sure that the AI system behaves like one coordinated organisation—not a set of disconnected bots. It checks that all decisions fit together, follow the rules, and support the company’s overall strategy before anything gets executed.

\section{Unified Agent System and Communication Logic}\label{sec:agent-equation}

To complete the agent-based system architecture, we introduce a unified optimisation structure that integrates both internal decision agents and human interface agents within a single analytical expression. This formulation provides a consistent basis for verifying system-wide admissibility, equilibrium, and institutional alignment. It enables the orchestration layer to assess the joint admissibility of all agent actions and user queries before any execution occurs. The integration of human-in-the-loop components into the agent architecture does not require the introduction of a separate communication agent. Instead, the dedicated interface agents—employee, stakeholder, and client—already serve as structured access points under role-based governance. These agents enforce strict admissibility rules through data filtering, role-tier mappings, and audit logs. Their outputs are routed through the System Orchestrator Agent, which validates whether all resulting actions and queries cohere with the institutional constraints of the firm. This routing logic preserves modular clarity while avoiding the control fragmentation that would result from parallel coordination layers. Introducing an additional communication core would dilute accountability, obscure institutional logic, and conflict with the compliance-by-design principles underlying the system. Formally, the global optimisation problem can be written as:
\begin{align}
\max_{\{\boldsymbol{a}_i\}, \{\boldsymbol{q}_j\}} \quad & \sum_{i=1}^{n} \mathbb{E}[U_i(\pi_i(\boldsymbol{a}_i, \boldsymbol{a}_{-i}))] + \sum_{j=1}^{m} \mathbb{E}[V_j(\mathcal{L}(\boldsymbol{q}_j, \boldsymbol{\alpha}_j))] \label{eq:global-agent-obj} \\
\text{s.t.} \quad
& \boldsymbol{a}_i \in \mathcal{D}_i \cap \mathcal{E}_i \cap \mathcal{G}_i \cap \mathcal{O}_i, \quad \forall i \label{eq:agent-admissibility} \\
& \boldsymbol{q}_j \in \mathcal{Q}_j, \quad \boldsymbol{\alpha}_j \subseteq \boldsymbol{A}_{\text{global}}, \quad r_j \in \mathcal{R}_{\text{permitted}}(j), \quad \forall j \label{eq:query-constraints} \\
& \{\boldsymbol{a}_i\}, \{\boldsymbol{q}_j\} \in \mathcal{A}_{\text{valid}} \quad \text{(system-level admissibility verified by orchestrator)} \label{eq:orchestrator-verification}
\end{align}

Here, $\boldsymbol{a}_i$ denotes the action vector of internal agent $i$ and $\pi_i(\cdot)$ is the agent-specific decision function, potentially dependent on the actions $\boldsymbol{a}_{-i}$ of other agents. The function $U_i(\cdot)$ measures the expected utility of the action, incorporating performance, compliance, or risk-based criteria. Each query $\boldsymbol{q}_j$ is submitted by a human-facing interface agent $j$ and evaluated through a large language model $\mathcal{L}(\cdot)$ applied to a filtered context $\boldsymbol{\alpha}_j$, drawn from the global agent state $\boldsymbol{A}_{\text{global}}$. The function $V_j(\cdot)$ captures the informational or operational utility returned to the user. The admissibility conditions restrict each internal action to the intersection of legally, environmentally, socially, and operationally permitted domains. Interface queries must be issued under valid role authorisations $r_j$, conform to data access constraints, and remain fully traceable. The orchestrator confirms that the combined set of actions and queries is admissible, non-contradictory, and institutionally consistent.

In simple terms, this equation brings all agent decisions—both machine-made and human-triggered—into one system-wide logic. Each agent tries to do its job well, and each user-facing agent responds only within allowed limits. Nothing proceeds until the orchestrator has checked that every action and answer fits together without breaking any legal or operational rule. This avoids having too many control layers and ensures the AI system behaves like one coherent institution.

\section{Protocols and Agent Integration in Enterprise AI}\label{sec:protocol}

The deployment of multiple agents in enterprise AI arises from the functional decomposition of insurance firms into distinct domains—capital allocation, underwriting, compliance, and claims handling—each governed by specific constraints, data privileges, and decision logic. Decentralisation alone is insufficient. To ensure coherence, regulatory admissibility, and shared utility optimisation, agents must be coordinated via protocol layers that standardise context propagation, state transfer, and inter-agent reasoning. This necessitates interoperability frameworks such as the Model Context Protocol (MCP) and Agent-to-Agent (A2A) messaging standards. Enterprise AI systems operate asynchronously across heterogeneous components with divergent compute schedules, partial observability, and uneven access to real-time data. Synchronous coordination introduces latency and fragility. Asynchronous architectures mitigate these issues by decoupling data availability from execution. Tadi \cite{Tadi2022} shows that asynchronous data handling enhances Progressive Web Application robustness, supporting offline operation and non-blocking updates—critical for agent continuity in intermittently connected environments. At the training level, asynchronous federated learning enables model updates without central synchronisation \cite{Zhao2024}, reinforcing system resilience. Beyond architectural theory, empirical studies demonstrate that agent communication latency increases with message size and agent count. Berna-Koes et al.\ \cite{berna-koes2004communication} show that message delays can grow non-linearly in multi-agent networks unless efficient backchannel protocols are implemented. Further, in delay-sensitive coordination tasks, latency-aware communication models such as DACOM \cite{yuan2022dacom} significantly enhance performance by adapting decision logic to inter-agent delays. Lei et al.\ \cite{lei2022latency} provide a framework for synchronising asynchronous perceptual features in collaborative systems, improving agent robustness in high-latency environments. In regulated domains such as insurance, empirical auditability standards further constrain system design. The UK Financial Conduct Authority requires AI systems to provide traceable decision logs within 24–72 hours of execution \cite{fca2022ai}. The Financial Stability Board emphasises the importance of explainability and audit resilience to manage systemic risks in AI-driven financial services \cite{fsb2017ai}. These empirical thresholds reinforce the need for persistent state management and protocol-level memory that can support post-hoc review under institutional supervision. The Agent-to-Agent (A2A) protocol, introduced by Google, standardises inter-agent messaging across frameworks such as LangGraph, CrewAI, and AutoGen by defining shared syntax for intention, memory, and task state \cite{GoogleA2A2024}. A2A allows agents built in distinct runtime environments to exchange structured content—decisions, delegation requests, alerts—while decoupling role semantics from implementation. In heterogeneous enterprise settings, where agents vary in logic paradigms or regulatory models, A2A enables coordination between e.g., an LLM-based compliance module and a rule-based pricing agent. As noted in \cite{AnalyticsVidhya2025}, A2A promotes composability by providing a shared abstraction for communication, aligning objectives under constrained interoperability. While A2A addresses syntactic and messaging compatibility, the Model Context Protocol (MCP) governs semantic coherence and memory persistence. MCP structures context propagation and state retention, enabling agents to reason over shared histories, hierarchical goals, and environmental signals \cite{Hou2024}. Krishnan \cite{Krishnan2025} highlights MCP’s capacity to encode long-range dependencies, allowing agents to retain mission-critical knowledge—e.g., regulatory thresholds, prior decisions, or user interactions—vital in financial or legal workflows. MCP functions as the institutional memory layer, necessary for consistency, justification of contingent decisions, and compliance transparency. Narajala \cite{Narajala2025} further underscores MCP’s role in auditability, enabling ESG alignment and AI Act compliance by preserving reasoning chains and allowing decision context replay. A2A and MCP are not substitutes but complements: A2A ensures technical interoperability and communication reliability; MCP secures semantic alignment and persistent memory. Together, they constitute a dual-stack foundation for robust, auditable multi-agent systems \cite{Hou2024, GoogleA2A2024}. MCP ensures that agent interactions are contextually meaningful; A2A ensures they remain operable across environments.

Protocol-level design is thus central to enterprise AI. The insurance architecture proposed here builds upon this dual-protocol paradigm, combining asynchronous operation with explainable, persistent decision memory to meet institutional and regulatory demands.

\section{Conclusion}\label{sec:conclusion}

This paper presents a formal architecture for implementing enterprise AI in regulated insurance firms, grounded in firm theory, constrained optimisation, and institutional logic. The insurer is modelled as a structured decision system subject to solvency constraints, legal admissibility, ESG obligations, and internal feasibility limits. This framework enables a principled decomposition of the firm’s behaviour into analytically distinct subfunctions aligned with regulatory and strategic demands.

A multi-agent architecture defines each functional domain—capital management, underwriting, claims, and compliance—as a bounded rational agent operating within a shared constraint environment. Local decision models integrate directly into the firm’s institutional structure, ensuring that optimisation occurs strictly within admissible legal, financial, and ESG boundaries.

An orchestrator agent enforces system-level coherence, validating inter-agent alignment, propagating global constraints, and maintaining institutional consistency. Protocol-level infrastructures such as asynchronous coordination, memory persistence, and semantic interoperability via MCP and A2A support structured communication across heterogeneous agent modules.

This framework advances the structural modelling of regulated firms by embedding regulatory, operational, and governance constraints into the decision logic itself. Unlike conventional models that treat compliance as an external layer, this architecture integrates institutional constraints into core decision processes, forming a foundation for building auditable, compliant AI systems in high-stakes institutional domains where formal accountability and systemic coherence are essential.

\bigskip

\bigskip

\printbibliography

\swDeclarationsPage

\end{document}